\documentclass[review]{elsarticle}
\usepackage[margin=2.5cm]{geometry}

\usepackage{graphicx}
\graphicspath{ {images/} }

\usepackage[section]{placeins}

\usepackage{lineno}
\usepackage{multirow}
\usepackage[colorlinks]{hyperref}

\modulolinenumbers[5]

\journal{Nuclear Instruments and Methods B}

\begin{document}

\begin{frontmatter}

\title{On The Relationship Between The Energy, Energy Spread\\And Distal Slope for Proton Therapy Observed in GEANT4}

\author[hudaddress]{Tim Fulcher\corref{correspondingauthor}}
\cortext[correspondingauthor]{Corresponding author}
\ead{timothy.fulcher@hud.ac.uk}
\author[ucladdress]{Richard A. Amos}
\author[cockaddress,manaddress]{Hywel Owen}
\author[stfcaddress,cernaddress]{Rob Edgecock}

\address[hudaddress]{University of Huddersfield, United Kingdom}
\address[cockaddress]{Cockcroft Institute of Accelerator Science and Technology}
\address[manaddress]{University of Manchester, United Kingdom}
\address[stfcaddress]{STFC, United Kingdom}
\address[cernaddress]{CERN}
\address[ucladdress]{Proton and Advanced Radiotherapy Group, Department of Medical Physics and Biomedical Engineering, University College London, United Kingdom}

\begin{abstract}
In proton therapy both the energy, which determines the range, and the distal slope, which reflects the rate at which the protons decelerate, are of import if we are to ensure accurate dose deposition and maximum tissue sparing. This publication describes a Geant4 model and presents a two-dimensional polynomial relationship between energy, the energy spread and the distal slope for beams with Gaussian energy spectra for proton therapy. This simple polynomial relationship will be useful for non-invasive or minimally invasive near real-time monitoring of the energy and energy spread of a proton therapy beam.
\end{abstract}


\end{frontmatter}


\section{Introduction}
A proton of a preset energy is possessed of a distinct or unique distance for a set material at which it, on average, releases its energy. The Bragg peak is, for many, counter intuitive. This mechanism is best described thus: a nucleus of the material through which the proton is passing has a proton capture cross section. The greater the proton capture cross-section the greater the probability that it will capture a proton. This proton capture cross-section increases with the decrease in energy as a result of slower particles experiencing the electric force of an atoms electrons for longer time. 

\pagebreak

The rate of loss of energy along the path of any proton may be statistically described by the Bethe-Bloch equation \cite{bethe_zur_1930} 

\begin{equation}
    -\frac{\mathrm{d}E}{\mathrm{d} s}=\frac{4\pi}{m_ec^2}.\frac{nz^2}{\beta^2}(\frac{e^2}{4\pi\epsilon_0}){ [ln(\frac{2m_ec^2\beta^2}{I(1-\beta^2)}-\beta^2]} 
\end{equation}

where:

\begin{center}
\begin{tabular}{ c c }
 $\beta$ & v/c \\ 
 v & velocity of the particle \\  
 E & energy of the particle \\  
 s & distance travelled by the particle \\
 c & speed of light \\
 z & particle charge \\
 e & charge of the electron \\
 $m_e$ & rest mass of the electron \\
 n & electron density of the target \\
 I & mean excitation potential of the target \\
 $\epsilon_0$ & permittivity of a vacuum
\end{tabular}
\end{center}

The range of a proton in a medium can be determined by integrating the stopping power from 0 to $E$
\begin{equation}
    R=\int_0^E-\frac{\mathrm{d}E}{\mathrm{d} s}\mathrm{d}E
\end{equation}

\begin{figure}[!h]
\begin{center}
   \includegraphics[scale=0.5]{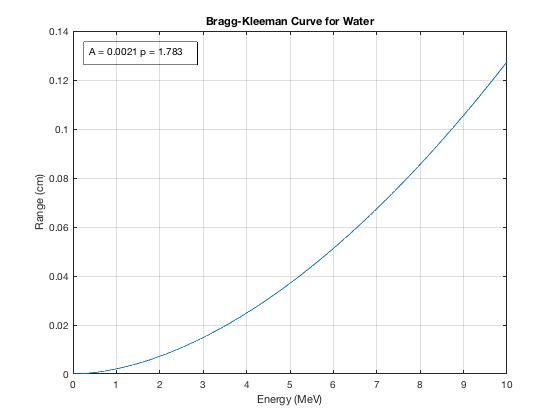}
\end{center}
\caption{The Bragg-Kleeman rule for water (A=0.0021, p=1.783).}
\label{fig:bk}
\end{figure}

This is known as the continuous slowing down approximation or CSDA \cite{boon_dosimetry_1998}. It is possible to determine the relationship between the CSDA range (in $g/cm^2$) and the initial energy by

\begin{equation}
    R=A x^p
\end{equation}

where $A$ and $p$ are constants for a particular medium, $x$ is the energy in MeV and $R$ is the depth in cm at which the Bragg peak, on average, manifests itself. This relationship is known as the Bragg-Kleeman rule \cite{bragg_ionisation_1904} and is shown in Figure \ref{fig:bk} for water.

\pagebreak

The importance of the correct beam energy being deposited at a position on a plane cannot be understated. Incorrect energy selection can result in a range change. This range change can result in healthy tissue receiving unwanted and, possibly, toxic dose. Figure 2 illustrates the difference in two Bragg Peaks of the same mean energy but with differing energy spreads.

\begin{figure}[!h]
\begin{center}$
\begin{array}{cc}
\includegraphics[scale=0.4]{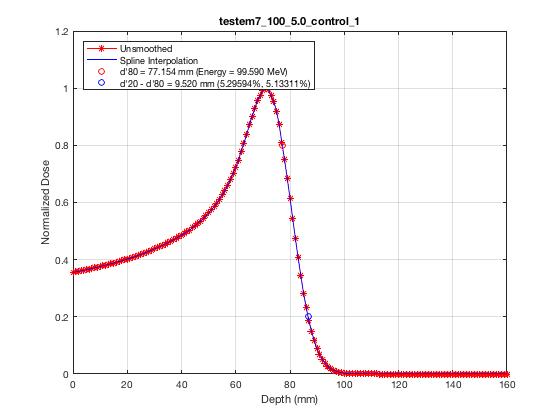}
\includegraphics[scale=0.4]{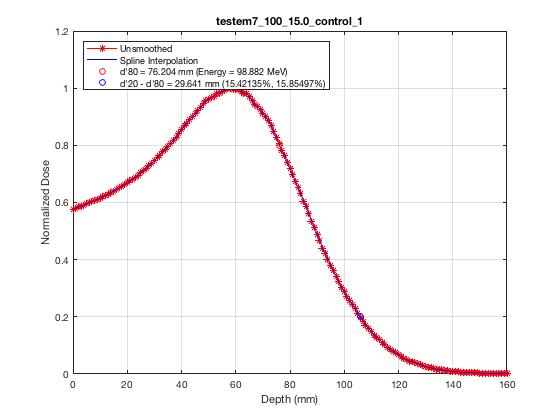}
\end{array}$
\end{center}
\caption{Two Bragg Peaks of the same energy but different energy spreads (left
= 5\%, right = 15\%). The difference in the distal slope should be apparent.}
\label{Fig:Race}
\end{figure}

Zalel et al \cite{zalel_analysis_2013}: “The energy spread of a proton beam causes a significant modification in shape of the Bragg Peak. This needs to be monitored and controlled in order to ensure the spread-out Bragg Peak or SOBP used in oncological treatments is modeled correctly so as to treat tumor efficiently and minimize damage to healthy tissue behind the tumor”. Additionally, Chu says: “There should be provided an instrumentation to measure the energy spread of the beams” \cite{chu_instrumentation_1993}. 

Maintaining beam characteristics for the clinical application of proton radiotherapy is of paramount importance. Treatment plans are designed using software that model the distribution of dose within a patients’ anatomy based on beam data measured during the commissioning of the proton delivery system. Variation of the distal penumbrae of Bragg peaks as a function of energy spread for a given nominal mean beam energy can lead to changes in the distribution of dose within the patient that may have undesirable clinical consequences. SOBP fields used for clinical treatments are the result of superposition of multiple “pristine” Bragg peaks are different ranges to cover the clinical target volume (CTV) within the patient. Variation of distal penumbrae of these Bragg peaks can lead to SOBP fields that are not optimal to ensure coverage of the CTV, and risk higher than planned dose to organs-at-risk (OAR) in close proximity to the CTV. Increasing dose to OAR can leads to radiation-related toxicities. Contemporary proton beam radiotherapy utilizes pencil beam scanning (PBS) technology to deliver highly complex intensity-modulated proton therapy (IMPT) plans. These rely of the accurate positioning of multiple “pristine” Bragg peaks within the CTV from numerous beam directions relative to the patient, and so the accuracy of the Bragg peak shapes as delivered compared to those calculated during treatment planning are vital in obtaining an optimal therapeutic and safe treatment. Moreover, interest is growing in being able to model the linear energy transfer (LET) variation within a clinical treatment plan to be able to better predict adverse biological reactions and to mitigate them. The LET has been shown to increase at the distal penumbra of a proton beam \cite{paganetti_relative_2014}, and so again being able to maintain the distal penumbra characteristics is important.  

It was thus desirable that a means of determining the energy and energy spread and a relationship between the two from a Bragg Peak be determined.

\section{Description of Modeling Methodology}
To determine a methodology, the energy loss of protons have been modelled using Geant4 \cite{geant4_collaboration_geant4_2003}. The modeling involved a modified version of the TestEM7 example provided with the Geant4 examples. The primary modification was the change from a mono-energetic source to a source with the capability of producing particles with energies with a Gaussian distribution. The pencil beam has a diameter of one sigma of 1 mm. A sample of 100,000 protons was used. Both momentum components transverse to the beam direction were set to 0.

A number of energies and Gaussian energy spreads deemed appropriate for proton therapy were entered into the Geant4 model. Those energies were 50, 70, 100, 150, 200 and 250 MeV and the energy spreads were 0.0, 0.1, 0.2, 0.5, 1.0, 2.0, 5.0, 10, 15, 20, 25, 30 and 35\%. All of one sigma or standard deviation of mean energy. Figure 3 shows a simple graphic representation of Geant4 model.

The medium was 1000 mm of water and 1001 steps were taken. The resultant data were processed in MATLAB [7] and a spline interpolation was used to interpolate the data.

\pagebreak

\begin{figure}[!h]
\begin{center}
   \includegraphics[scale=0.5]{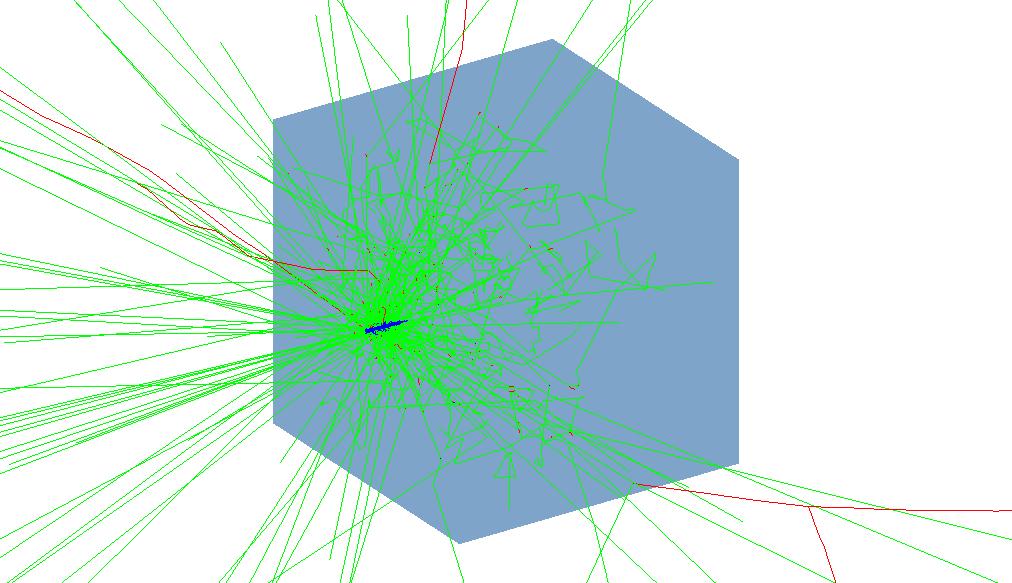}
\end{center}
\caption{BDSIM \cite{laurence_nevay_bdsim_2018} representation of the Geant4 model of the water phantom. The beam is incident to the left face of the water phantom cube (in light blue).}
\label{fig:BD_SIM_Model}
\end{figure}

When determining the distal slope numerous points on the distal slope have been mooted \cite{gottschalk_characterization_2003}, one of which is the peak to d’80 or 80\% of the normalized dose on the distal slope. The d’80 point is also used to determine the range. However the peak to d’80 point as a metric is far from unique. Another commonly used metric is to subtract the d’80 point from the d’20 point of the normalized dose. This is referred to as the d’20 to d’80 distance. A benefit of this metric is that it is independent of the peak value and the vagaries of the proximal region.

\subsection{Straggling}
Before discussing the model we should consider the effects of straggling. Bohr \cite{bohr_decrease_1915} demonstrated that for long path lengths the distribution of the stopping distance about the mean is approximately Gaussian. Chu \cite{chu_instrumentation_1993} reported a convenient power law approximation to estimate sigma as a function of the proton beam range.

\begin{equation}
    \sigma_\delta=kR_0^m
\end{equation}

where R\textsubscript{0} is the range in water in centimeters for a mono-energetic proton beam, k is a material-dependent constant of proportionality, and the exponent is empirically determined. For protons in water, k = 0.012 and m = 0.935 \cite{bortfeld_analytical_1997}.

\begin{figure}[!h]
\begin{center}
   \includegraphics[scale=0.5]{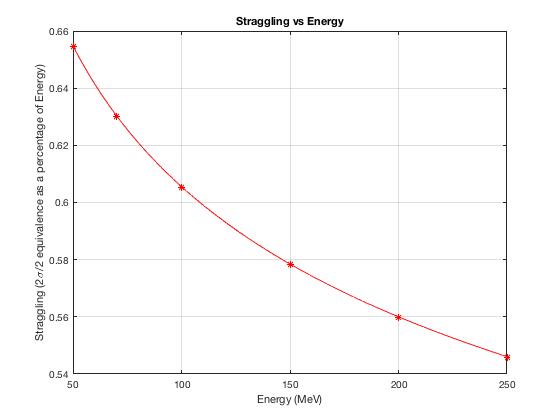}
\end{center}
\caption{Range straggling in water equivalence as a function of energy. The peculiar vertical scale is as a result of the asymmetrical nature of the calculated energy spread when plotted about the mean.}
\label{fig:straggling}
\end{figure}

\pagebreak

In a perfect medium, one which is free of straggling, figure \ref{fig:straggling} represents the energy spread that would be needed to be introduced to observe the equivalent resultant from straggling observed in a normal or imperfect medium. Clearly the range straggling as a percentage of energy is energy or range dependent and decreases as a function of energy as that energy or range increases.

Manipulation of equation 4 results in the range straggling equivalence as a function of energy and we obtain the relationship illustrated in figure 4. 

The straggling effects are important because they result in the flattening of the lower end of the curve, resulting in a ‘hockey stick’ effect. These effects may be considered  analogous to a signal emerging from the noise level. Zalel et al \cite{zalel_analysis_2013} observed a similar flattening of the lower end of the curve.


\section{Development of Model}
The methodology described in section 2 was used to model the SOBP for various energies and energy spreads described in that section.

Using MATLAB, the resultant d’20 – d’80 distances were calculated for the previously listed energies and energy spreads. Figure 5. shows the resultant curves.

\begin{figure}[!h]
\begin{center}
   \includegraphics[scale=0.5]{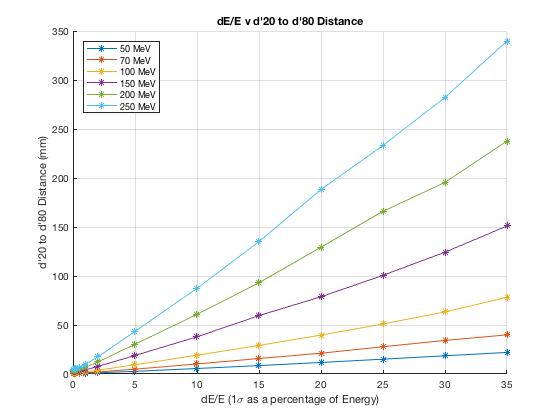}
\end{center}
\caption{Curves describing the relationship between a Gaussian energy spread and the d’20 to d’80 value of a Bragg-Peak generated by differing energies and energy spreads. The effects of straggling become noticeable below 1\% as a flattening out of the data (the "hockey stick" effect mentioned in Section 2.1).}
\label{fig:raw_d20_d80_curves}
\end{figure}

\pagebreak

\begin{figure}[!h]
\begin{center}
   \includegraphics[scale=0.5]{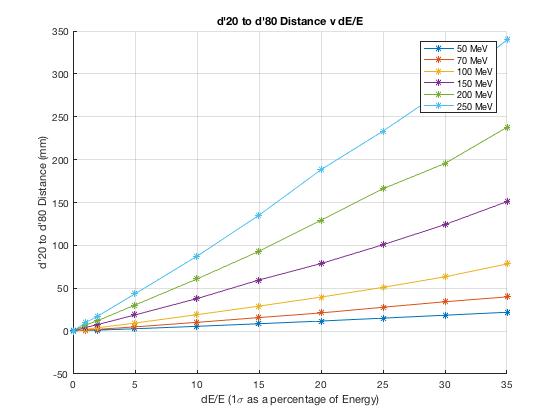}
\end{center}
\caption{Figure \ref{fig:raw_d20_d80_curves}. with the 0.0, 0.1, 0.2 and 0.5 dE/E values omitted and with the y-intercept resultant from the curve fitting included.}
\label{fig:d20_d80_curves_no_hockeystick}
\end{figure}

Readings below 1.0\% were omitted because the fitting of those curves would be complicated by the effects of range straggling. The 0\% point was interpolated by using the MATLAB basic curve fitting tool and incorporated the y-intercepts for each energy. The resultant plots are show in figure \ref{fig:d20_d80_curves_no_hockeystick}

\pagebreak

\begin{figure}[!h]
\begin{center}
   \includegraphics[scale=0.5]{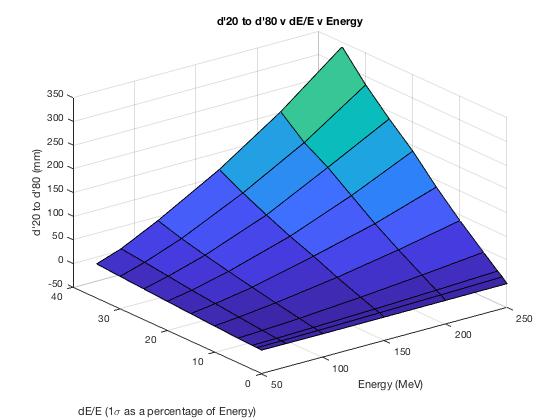}
\end{center}
\caption{Figure. 6. expressed as a surface plot.}
\label{fig:surface}
\end{figure}

Figure \ref{fig:d20_d80_curves_no_hockeystick}. was then plotted as a surface plot yielding figure \ref{fig:surface}.

The data in figure 7 was processed using the MATLAB curve fitting toolbox and the polynomial described in equation 5 was determined to be the best fit yielding an R squared value of 0.9998.

\begin{equation}
    z=4.394-0.1078x-0.5001y+0.0007811x^2+0.01413xy-0.001926y^2-1.668e-6x^3-8.816e-5x^2y-0.0001366xy^2
\end{equation}

Where z is the d’20 to d’80 distance in millimeters, x is the Mean Energy in MeV and y is the energy spread as a percentage of mean energy. 

\pagebreak

Thereafter the polynomial described in equation 5 was plotted yielding figure 8 

\begin{figure}[!h]
\begin{center}
   \includegraphics[scale=0.5]{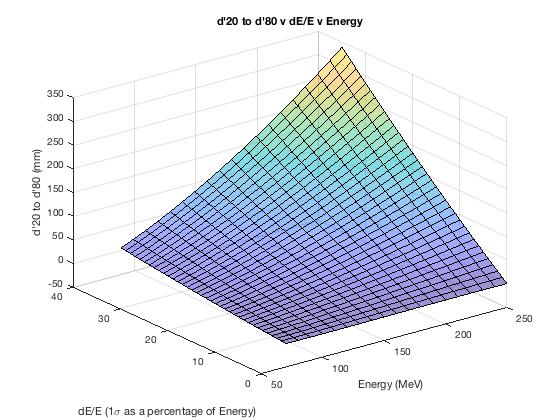}
\end{center}
\caption{Interpolated surface plot resultant from equation 5.}
\label{fig:interpolated_surface}
\end{figure}

\begin{figure}[!h]
\begin{center}
   \includegraphics[scale=0.5]{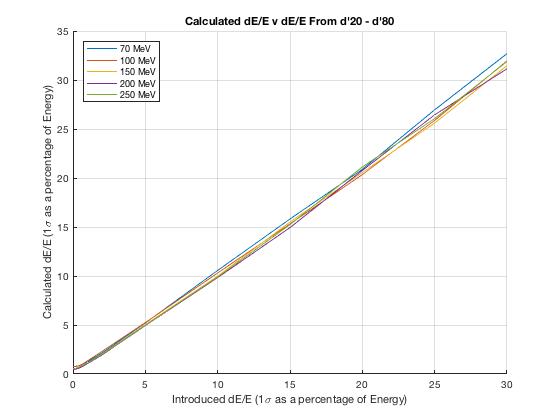}
\end{center}
\caption{Plots of the response of the polynomial model for various energies
and energy spreads of up to 30\%. Evidence of the slight flattening tendencies at lower energies, might be ascribed to the straggling effect
described previously. A noticeable deviation from the linear is apparent above energy spreads of 20\%.}
\label{fig:response_all_energies}
\end{figure}

Energy spreads for several energies were introduced and plotted against that which was calculated for that energy using the derived polynomial model until a clear non-linearity was observed (see figure. \ref{fig:response_all_energies}).

\section{Validation and Results}
\begin{figure}[!h]
\begin{center}
   \includegraphics[scale=0.5]{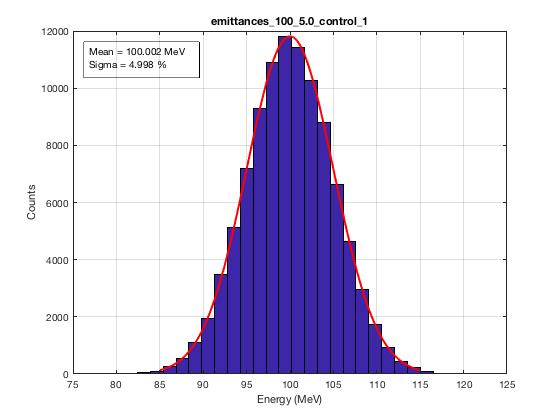}
\end{center}
\caption{An energy spectrum resultant from the modified Geant4 code
for an introduced beam with mu = 100.0 MeV and sigma of 5\%. The
Gaussian fit returns mu = 100.002 MeV and sigma = 4.998\%.}
\label{fig:energy_spectrum}
\end{figure}

In order to facilitate the validation of the polynomial model the Geant4 code was further modified such that the energies of the protons were gauged as they entered the water phantom and these values were imported into MATLAB and plotted in a histogram and a Gaussian fit applied to that histogram an example of which is illustrated in figure \ref{fig:energy_spectrum}.

\pagebreak

\begin{figure}[!htb]
\begin{center}
$\begin{array}{cc}
\includegraphics[scale=0.4]{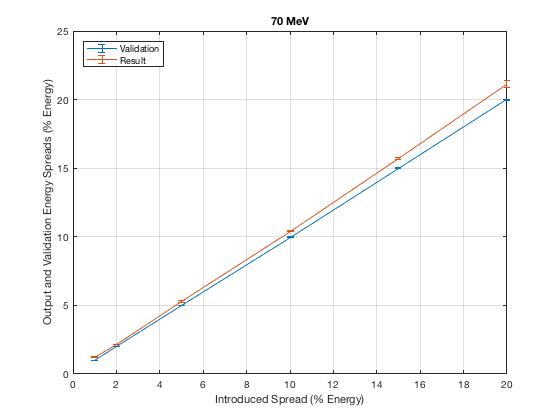}
\includegraphics[scale=0.4]{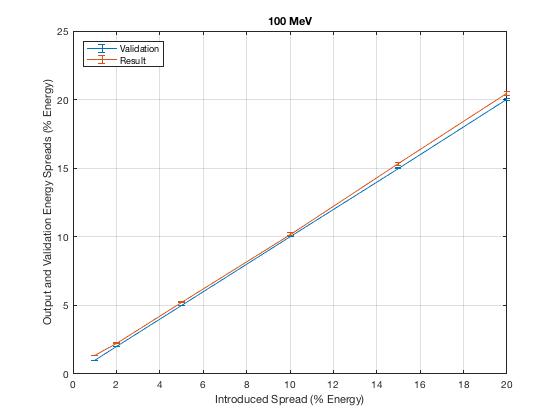}
\end{array}$
$\begin{array}{cc}
\includegraphics[scale=0.4]{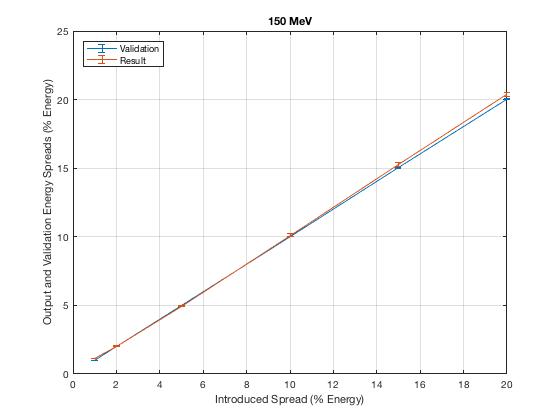}
\includegraphics[scale=0.4]{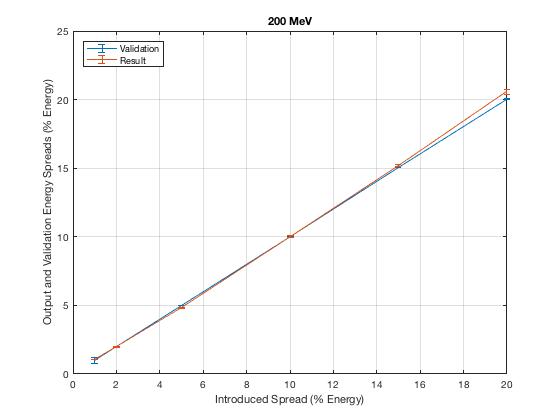}
\end{array}$
$\begin{array}{cc}
\includegraphics[scale=0.4]{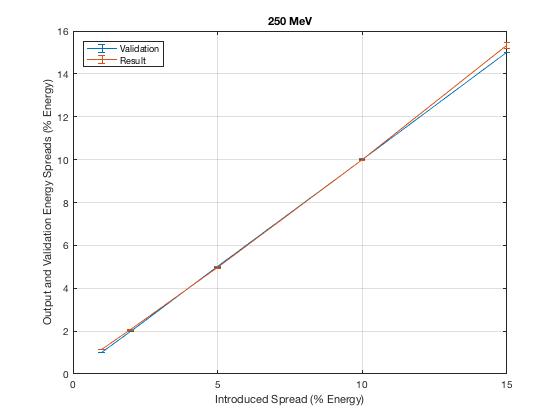}
\end{array}$
\end{center}
\caption{Plots of the input versus output response of the validation and
the polynomial model for several introduced energies. The final plot of
20\% energy spread at 250 MeV is omitted because of a recurrent
segmentation fault in Geant4 at that energy spread}
\label{fig:response_v_validation}
\end{figure}

The Gaussian values (mu and sigma, the determination of which was described above) were compared to those introduced to and derived from the polynomial model (equation. 5) and the following relationships were observed within the 1\% - 20\% range (figure \ref{fig:response_v_validation}).

\pagebreak

\begin{figure}[!htb]
\begin{center}
$\begin{array}{cc}
\includegraphics[scale=0.4]{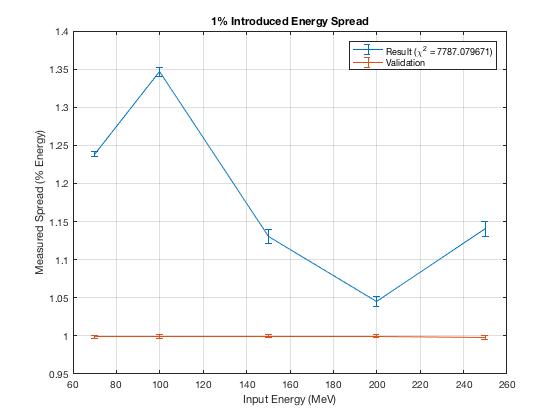}
\includegraphics[scale=0.4]{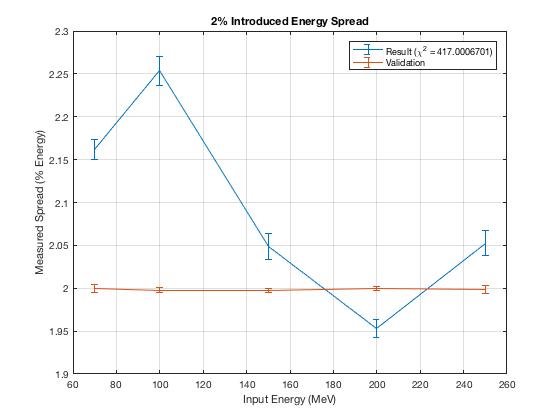}
\end{array}$
$\begin{array}{cc}
\includegraphics[scale=0.4]{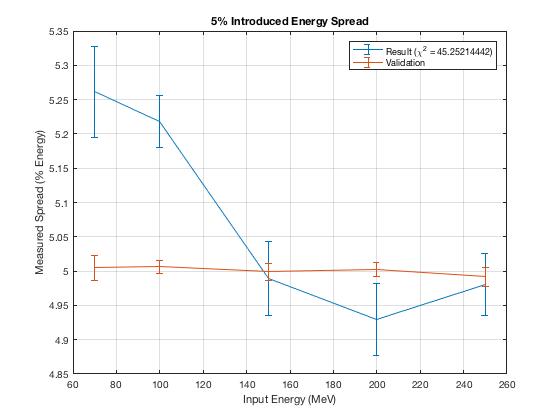}
\includegraphics[scale=0.4]{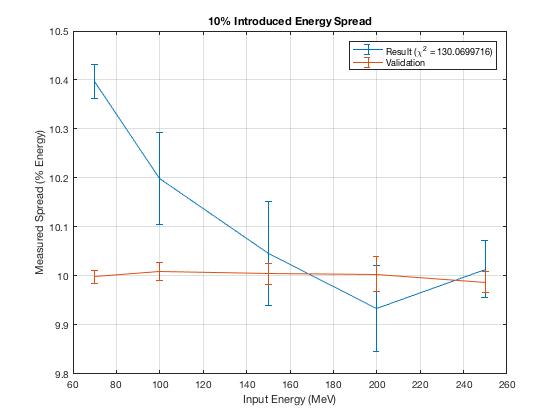}
\end{array}$
$\begin{array}{cc}
\includegraphics[scale=0.4]{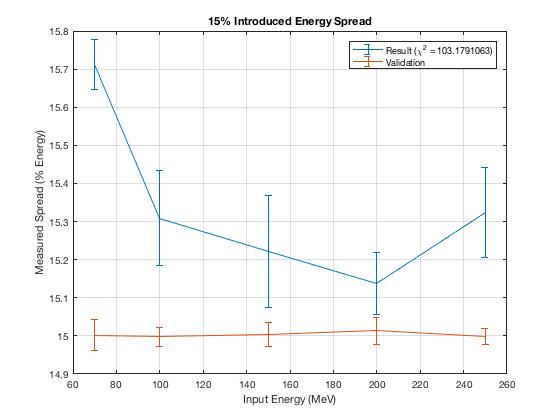}
\includegraphics[scale=0.4]{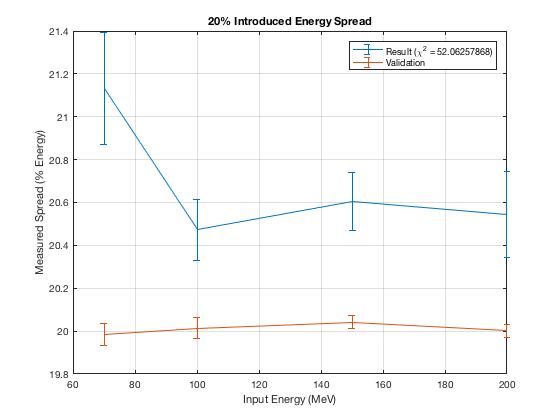}
\end{array}$
\end{center}
\caption{Plots of the input versus output response of the validation and the
polynomial model for several introduced energy spreads. Again, the final plot
of 20\% energy spread at 250 MeV is omitted because of a recurrent
segmentation fault in Geant4 at that energy spread.}
\label{fig:spreads_v_energy}
\end{figure}

By plotting the resultant energy spreads by energy we obtain
the following graphs (figure \ref{fig:spreads_v_energy}).

\pagebreak

The 70 MeV curve in figure \ref{fig:spreads_v_energy} is, clearly, outside values that can be reasonably expected, the remainder of the energy response curves exhibited in figure \ref{fig:spreads_v_energy} show a close correlation between the stimulus and response. While the 1\% response in figure figure \ref{fig:spreads_v_energy}  is outside the parameters expected for a medical device the remainder of the curves are exhibit responses that are close to and within those parameters.

\begin{figure}[!h]
\begin{center}
$\begin{array}{cc}
\includegraphics[scale=0.31]{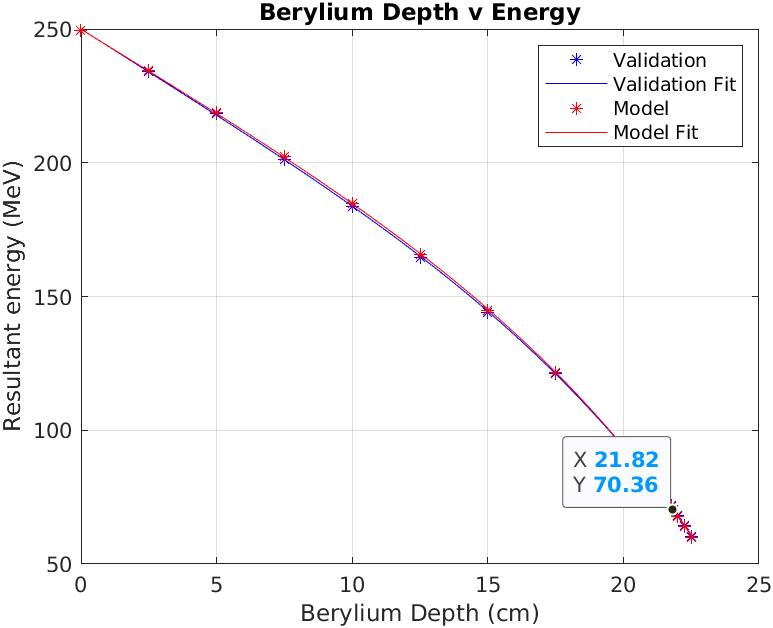}
\includegraphics[scale=0.31]{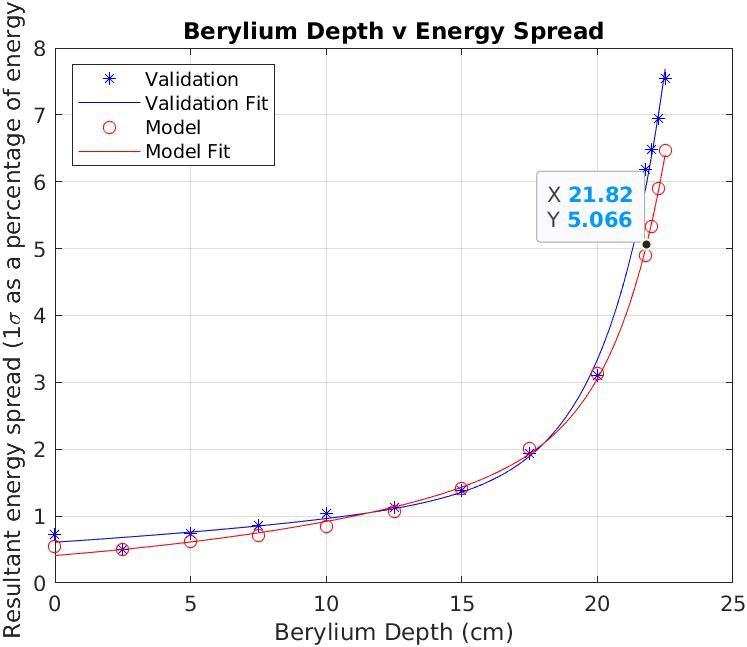}
\end{array}$
\end{center}
\caption{Plots of energy degradation through Beryllium for
various thicknesses thereof (left) and the
accompanying increase in energy spread (right). The
energy spread of 5.066\% at 70.36 MeV agrees well
with Liang and van Goethem}
\label{Fig:Beryllium}
\end{figure}

Van Goetham et al \cite{van_goethem_geant4_2009} and Liang et al \cite{zhikai_liang_design_nodate} have cited an energy
spread of approximately 4.9 \% in Beryllium (density = 1.9 g/cm3 and
ionization potential of 63.7 eV) with a resultant energy of
approximately 70 MeV and an incident energy of 250 MeV and no
incident energy spread. From figure \ref{Fig:Beryllium} it can be seen that with similar
input values and by examining the Bragg Peak the polynomial model
returns a very similar energy spread in Beryllium.

Similar to the validation method described earlier the spectrum of the beam emergant from the Berylium degrader was captured and used as a validation for results derived from the water phantom model.

\pagebreak


\section{Conclusions}
The polynomial in equation 5. has been found to yield results that are near to if not within clinical tolerances between, approximately, one sigma of 1\% and 15\%  energy spread for an environment modelled in Geant4 for energies between 100 MeV and 250 MeV. We believe this model may be useful in determining the energy spread of a therapeutic proton therapy beam after further refinement.

If the energy spread can be monitored real time during beam extraction, then the relationship between energy spread and distal penumbra can be calculated and compared to that used for the clinical beam model. With an appropriate feedback system in place, the beam characteristics may either be changed, or the beam terminated during clinical delivery. Clinical tolerances are typically 2\%. The model presented here shows good agreement between introduced and modelled energy spreads for approximately 5-15\% spread for 120-250 MeV proton beams. These will enable the model to be used for many treat scenarios, but not all. Shallow tumours require proton beam energies of between 70 -100 MeV, and so further development of the model for lower energy beams is required. 

It would be of interest to see whether this or any subsequent refinements thereof modelled relationship could be observed in a water phantom.

\pagebreak

\bibliography{My_Library_VII.bib}

\end{document}